\documentclass[aps,prl,twocolumn,showpacs,superscriptaddress,groupedaddress,longbibliography]{revtex4-1}  
\usepackage{graphicx}  
\usepackage{dcolumn}   
\usepackage{bm}        
\usepackage{amssymb}   

\pdfoutput=1
\usepackage{color}
\usepackage{subfigure}
\usepackage{lineno}

\usepackage{latexsym,amsmath,verbatim}

\begin{document}

\title{Avalanche precursors of failure in hierarchical fuse networks}
\author{Paolo Moretti, Bastien Dietemann}
\affiliation{Dept. of Materials Science, WW8-Materials Simulation, FAU Universit\"at Erlangen-N\"urnberg, Dr.-Mack-Stra{\ss}e 77, 90762 F\"urth, Germany}
\author{Michael Zaiser}
\affiliation{Dept. of Materials Science, WW8-Materials Simulation, FAU Universit\"at Erlangen-N\"urnberg, Dr.-Mack-Stra{\ss}e 77, 90762 F\"urth, Germany}
\affiliation{
	School of Mechanics and Engineering, Southwest Jiaotong University, Chengdu 610031, China
}

\begin{abstract} 
We study precursors of failure in hierarchical random fuse network models which can be considered as idealizations of hierarchical 
(bio)materials where fibrous assemblies are held together by multi-level (hierarchical) cross-links. When such structures are loaded towards failure, the patterns of precursory avalanche activity exhibit generic scale invariance: Irrespective of load, precursor activity is characterized by power-law avalanche size distributions without apparent cut-off, with power-law exponents that decrease continuously with increasing load. This failure behavior and the ensuing super-rough crack morphology differ significantly from the findings in non-hierarchical structures.    
\end{abstract}


\maketitle

Hierarchical materials are characterized by microstructure features that repeat on different length scales in a self-similar fashion. Biological materials provide compelling examples. Collagen, for instance, exhibits a hierarchical fiber organization which at different length scales comprises molecules, microfibrils, fibers, and fiber bundles \cite{Gautieri2011_NL}. Such complex organization was shown to provide enhanced toughness over assemblies of isolated collagen molecules. Several authors (see e.g. \cite{Gao2006_IJF}) have suggested that hierarchical organization may delay or prevent the nucleation and spreading of critical flaws which control failure of non-hierarchical heterogeneous materials \cite{Lennartz2013_PRE,Zaiser2015_JSTAT}. Models of hierarchical materials have mostly used hierarchical generalizations of the well-known equal-load-sharing fiber bundle model (ELS-FBM) which is a mean-field model for brittle fracture in disordered materials (see e.g. \cite{Zapperi1999_PRE}). In hierarchical variants, fibers are recursively grouped into bundles and load is assumed to be distributed equally among the intact fibers within each bundle - a salient feature which makes such models amenable to analytical treatment as renormalization arguments can be used to deduce the overall strength \cite{Newman1994_Physica} and the statistics
of damage accumulation. Hierarchical fiber bundle models have been used in the context of biomaterials (see e.g. \cite{Pugno2012_PRE}) and also of composites \cite{Mishnaevsky2011_CST}. A variant which consists in envisaging the structural elements of a hierarchical fiber bundle not as simple fibers but as chains-of-bundles does not greatly alter the basic conceptual framework since, at least in the limit of elastic-brittle local constitutive behavior, the properties of a bundle can be inferred from those of the single fibers using standard methods \cite{Pradhan2010_RMP} and those of a chain-of-bundles then be deduced by weakest-link statistics. Models of this type were introduced for a speculative nanotube-space-elevator cable \cite{Pugno2008_small} and for hierarchical bio-materials \cite{Pugno2012_PRE}. 

Practically all investigations of hierarchical fiber bundles focus on the effective strength of the hierarchical structures, whereas fundamental questions concerning the {\em nature} of the failure process (critical behavior vs. sub-critical crack nucleation-and-growth) and the concomitant nature and statistics of precursor events have received little attention \cite{Pradhan2010_RMP}. In fact, because of their mean-field nature, ELS-FBM and their generalizations are not well suited for the investigation spatial patterns of damage accumulation and failure. In the present work we therefore depart from the fiber bundle paradigm. To investigate how hierarchical organization affects the precursor activity in the run-up to failure and ultimately changes the {\em mode} of failure, we formulate for the first time hierarchical generalizations of the well-known random fuse network (RFN) \cite{deArcangelis1985_JPL,Alava2006_AP} which, unlike ELS-FBM, is known to capture esssential features of spatial stress patterns occurring during failure of continuous media such as crack-tip stress singularities.

\begin{figure*}
\includegraphics[width=14cm]{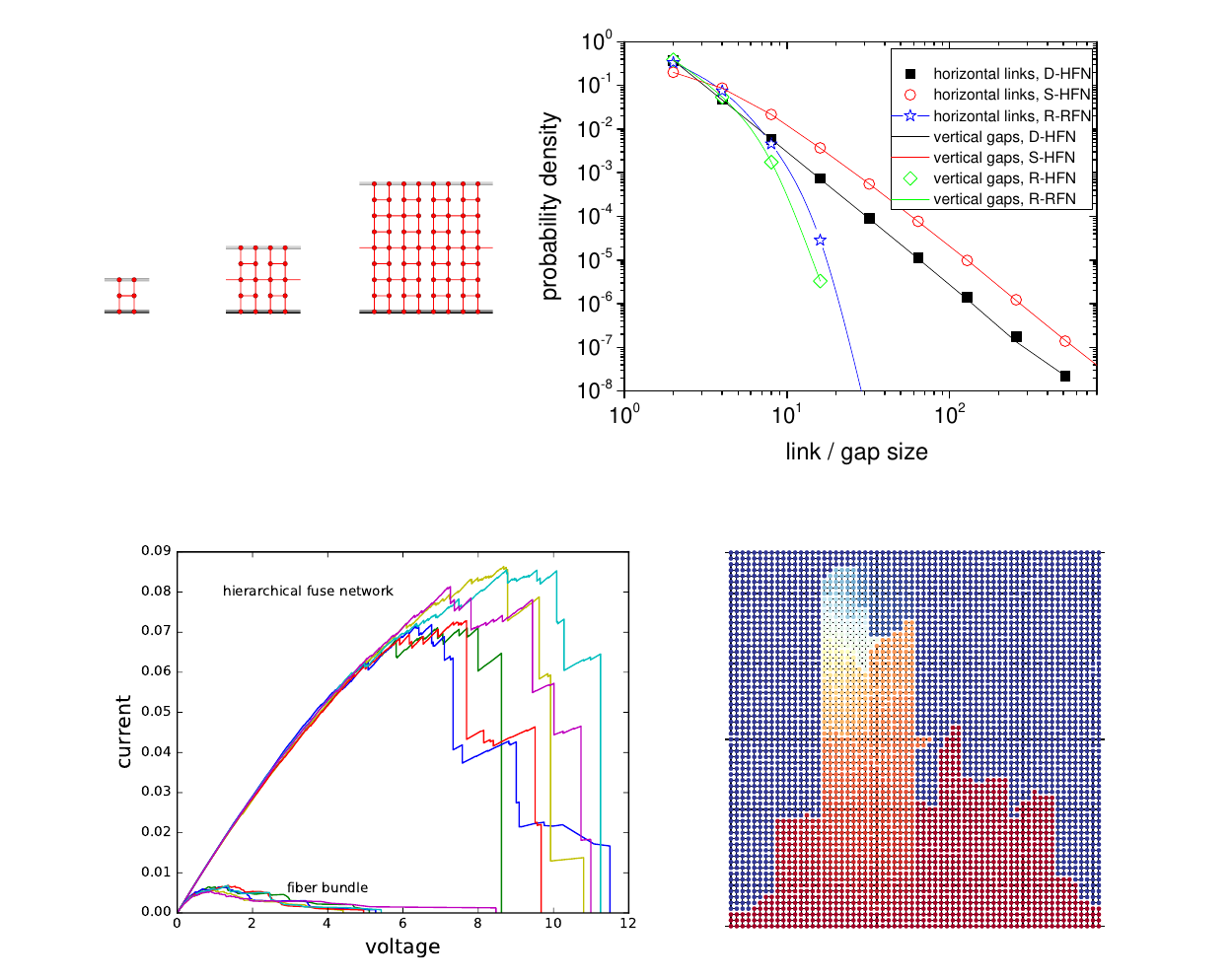}
\caption{Top left: Structure of the hierarchical fuse network, visualized as a recursive construction process, shown are the first three recursive steps; Top right: statistics of load-perpendicular cross links and load-parallel 'gaps' for regular and randomized HFN: regular D-HFN, S-RFN with random permutation of both columns and rows, R-HFN with randomly rotated rows, R-RFN reference random fuse model with randomly placed cross links: Bottom left: current-voltage curves for a set of S-HFN realizations with $L=64$ together with a set of realizations of reference fiber bundles with the same total number of links; bottom right: crack shape in a HFN with $L=64$ just before failure.}
\label{fig:HFN}
\end{figure*}

Our aim is to investigate the impact of hierarchical architecture on the mode of failure, to highlight substantial differences from non-hierarchical materials, and to draw analogies with the behavior of hierarchically architectured systems outside the realm of materials mechanics. To this end, we generalize the RFN model into a hierarchically cross-linked network of breakable fibers of heterogeneously distributed strength, which we denote as Hierarchcial Fuse Network (HFN). We consider a two-dimensional network of $N = L \times L$ nodes $k$ connected by bonds representing fuses of unit conductance. The network is contained between two bus bars, with periodic boundary conditions applied in the direction parallel to the buses. Loading is performed by imposing a current or a voltage between the bus bars. Bonds are arranged in such a manner that they form $L$ continuous fibers of length $L + 1$ connecting the bus bars. These fibers are mutually cross-linked in a deterministic, hierarchical manner that can be best understood in terms of a recursive construction as shown in Fig. \ref{fig:HFN}, top left: The network comprises smaller modules, which are recursively paired into larger ones by establishing rarer and longer cross-links. As a consequence of this construction, cross-links of exponentially increasing length are found with exponentially decreasing probability, and the same is true for the gaps that exist parallel to the loading direction. As a consequence, both cross-link lengths $n_{\rm cl}$ (number of horizontally connected cross links) and gap lengths $n_{\rm gp}$ (number of vertically adjacent gaps) are power-law distributed, $p(n_{\rm gp}) \propto p(n_{\rm cl}) \propto n_{\rm gp,cl}^{-\kappa}$ with $\kappa = 3$ for the construction shown in Fig. \ref{fig:HFN} (black squares and connecting black line in Figure \ref{fig:HFN}, top right). Other exponents can be implemented by altering the numbers of load carrying links and of cross links in the elementary module. The total number of cross linking bonds is denoted as $E$ with $E \approx 2/3 L^2$. The resulting deterministically constructed HFN will in the following be denoted as D-HFN. 

Randomized versions of the same type of network are constructed in several manners. (i) A network constructed by starting from a D-RFN and then first randomly reshuffling the columns and then the resulting rows is denoted as S-HFN (first reshuffling the rows and then the columns produces statistically equivalent results). Like D-HFN, S-HFN exhibit identical power-law distributions of cross-link lengths and gap lengths. While the short-length behavior of the distributions is slightly modified as compared to D-HFN, the power-law exponent of the distributions which governs the decay at large scales is unaltered (red circles and connecting red line in Figure \ref{fig:HFN}, top right), (ii) A network 
constructed by independently shifting the rows of a D-HFN by random fractions of $L$ across the periodic boundaries is denoted as R-HFN. The cross-link lengths of R-HFN are by construction identical to those of the initial D-HFN but the distribution of vertical gap sizes becomes  exponential, $p_{\rm gp} \propto \exp(-n_{\rm gp}/\bar{n}_{\rm gp})$ (iii) We take the $E$ cross-links and distribute them randomly over the $L^2$ possible cross-linking sites. This is denoted as a reference random fuse network, R-RFN. The R-RFN has the same exponential distribution of vertical gap sizes as the R-HFN but now also the distribution of cross-link lengths is exponential. Cross-link and vertical gap distributions for networks of size $L=1024$ are shown in Fig. \ref{fig:HFN}, top right. 

The networks are loaded by adjusting the voltage difference $V$ between the bus bars to maintain a fixed total current $I$ (load control). The voltage $V_k$ at node $k$ represents a displacement-like variable, while the currents $i_{kl}$ flowing between nodes represent stress-like variables. The equilibrium equations for this scalar model of elasticity result from Kirchhoff's law, imposing that the algebraic sum of all forces (currents) at a node must be zero. A bond connecting nodes $k,l$ fails irreversibly once the local current $i_{kl}$ exceeds a critical value $t_{kl}$. Stochastic material heterogeneity is mimicked by taking these thresholds to be independent random variables which we assume to be uniformly distributed between $0$ and $1$, representing an assembly of highly unreliable elements. We follow the standard  loading protocol for quasi-static RFN simulations \cite{Alava2006_AP}. The external load (the imposed current) is increased to the precise level where the first bond breaks and then kept fixed while bond failure leads to load re-distribution which may trigger further failures: damage accumulates through bursts of local failures (avalanches). The number of failures occurring as a consequence of internal load re-distribution at fixed total current defines the avalanche size $s$. Subsequent to an avalanche the load is again increased to induce bond breaking, and this is repeated until global failure disconnects the network.

Fig. \ref{fig:HFN}, bottom left, shows typical current-voltage characteristics for the HFN. For illustration we compare the performance of the simulated HFNs with reference fiber bundles constructed as follows: we remove all cross links from a HFN of a given size $L$ and use the $E$ removed bonds to create additional load bearing chains of length $L+1$ in a reference fiber bundle (we add a small number of additional bonds if $E$ is not a multiple of $L+1$). In the HFN, the cross links are initially load free, hence the load per longitudinal bond in the reference fiber bundle is initially less than in the HFN. Nevertheless the HFN out-performs the reference fiber bundle by a huge margin, yielding for $L+64$ a mean failure current that is more than one order of magnitude higher and a fracture energy (area under the $I-V$ curve) that is almost two orders of magnitude higher. For larger $L$ these discrepancies increase in approximate proportion with $L$ for the peak current, and with $L^2$ for the fracture energy. We finally show in Fig. \ref{fig:HFN}, bottom right, a typical crack profile for a HFN close to failure. 
Colors indicate different voltage values: a color discontinuity signals absence of conductance, due to a crack. The wide jumps in the crack profile are imposed by the underlying hierarchical structure, and are reminiscent of super-rough crack profiles as encountered e.g. in bone \cite{Launey2010_ARMR}. We note that this crack morphology strongly differs from the observations in non-hierarchical random fuse networks where generally compact, self affine crack shapes are observed \cite{Zapperi2005_PRE}.

\begin{figure*}
\begin{center}
\vspace{-0cm}
\includegraphics[width=0.8\textwidth]{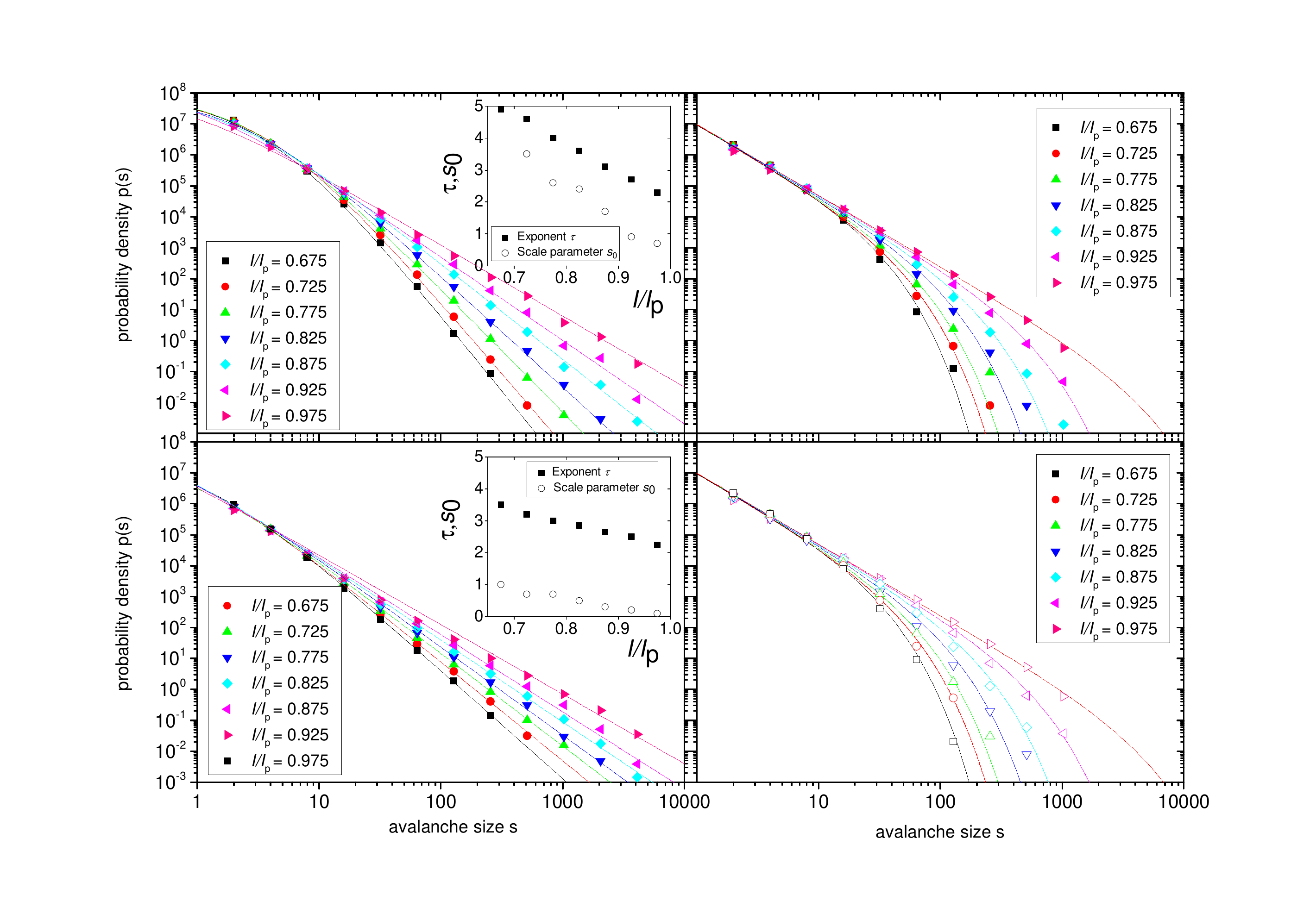}
\caption{Avalanche size distributions for HFN and random reference networks of size $L=512$; top left: D-HFN, bottom left: S-HFN, the lines represent fits of Pareto distributions as given by Eq. (2), the fit parameters are shown in the insets; top right: R-RFN, bottom right: R-HFN, the lines represent a common fit to all data (R-RFN and R-HFN) using Eq. (1); all distributions are averaged over $3\times 10^5$ realizations of the respective networks.}
\label{fig:avalanchestat}
\end{center}
\end{figure*}

We now study the size distributions of avalanches of bond breakings that occur prior to global system failure. We resolve these distributions with respect to the applied load (current): the loading curve is subdivided into load value intervals and avalanche size distributions are computed separately for each interval. For non-hierarchical RFN, the statistics of precursors to failure is well established: Avalanche activity in the run-up to failure is characterized by truncated power-law distributions of avalanche sizes of the form
\begin{align}
P(s) = N s^{-\tau} \exp\left[-\frac{s}{s_0}\left(1-\frac{I}{I_{\rm p}}\right)^{1/\sigma}\right]
\label{eq:truncpower}
\end{align}
with a fixed exponent $\tau$ and a cut-off that increases with load and diverges at the point of failure \cite{Zapperi2005_PRE,Zapperi2005_PA}. More realistic spring or beam models \cite{Nukala2005_PRE,Kun2014_PRL} yield similar results. The same picture can also be found in our own simulations of R-RFN where the lateral cross-links between the load carrying fibers are located randomly to create a non-hierarchical reference structure,  see Figure \ref{fig:avalanchestat}, top right, where the avalanche size distributions can be well fitted by Eq. \ref{eq:truncpower} with $\tau = 2.3$ and $1/\sigma = 1.95$. For comparison, Ref. \cite{Zapperi2005_PRE} reports values of $\tau  \approx 2$ and $1/\sigma \approx 1.4$ with a weak dependence on lattice morphology. While these exponent values differ from the mean-field values $\tau = 1.5$ and $\sigma=1$, the avalanche size distributions are of the same type as for ELS-FBM. 

In the case of HFN, the picture is completely different as power laws with continuously varying exponents are observed throughout the loading curve without an apparent cut-off. The distributions cannot be meaningfully be fitted by Eq. \ref{eq:truncpower} but are well represented by modified Pareto distributions, 
\begin{align}
P(s)= \frac{N}{s+s_0}^{-\tau} 
\label{eq:Pareto}
\end{align}
where now the exponent $\tau$ decreases with increasing load $I$ in an approximately linear manner (Figure \ref{fig:avalanchestat}, left). Only at the peak current the distributions for HFN and RFN approach each other, as in the former case the cut-off diverges while for the HFN the exponent of the scale free distribution approaches the asymptotic value $\tau = 2.3$ that also characterizes the random reference network. We may thus conclude that, whereas RFN exhibit a kind of critical-like behavior which is scale free only at the point of failure, in  HFN such scale free behavior is a robust, intrinsic features of the dynamics as the avalanche size distributions have power-law characteristics without cut-off even far away from the peak load. 

In order to understand the origin of this robust scale free behavior, we note that hierarchical modular organization has been known to produce generic scale invariant behavior in systems apparrently unrelated to materials mechanics. Models of activity propagation in both real and computer generated mappings of the human brain, in particular, have produced similar avalanche size distributions with continuously varying, non-universal exponents \cite{Moretti2013}. Power-law distributed avalanche sizes are believed to be a direct consequence of the morphology  of the brain networks, which are organized into a hierarchy of modules of exponentially increasing size yet exponentially decreasing number. Thus, scale free dynamic patterns are a consequence of scale-free hierarchical organization of the underlying network, a consideration that holds for processes as varied as activity propagation and percolation, and is backed by renormalization results \cite{Friedman2013}. 

The HFN architecture considered in the present work gives rise to scale free features on two distinct levels, as both the length distribution of transversal cross-links connecting the load-carrying fibers and the length distribution of longitudinal 'gaps' separating fibers are characterized by power laws. To understand the role of these features in ensuring the scale free statistics of precursor activity, we compare the behavior of the different network variants. The behavior of the D-HFN and the randomly re-shuffled S-HFN is essentially the same: in both cases we observe power-law avalanche size distributions with an exponent $\tau$ that decreases towards the value at failure, $\tau = 2.2$, as an approximately linear function of the current $I$. At large avalanche sizes, the distributions are very clean power laws, at small sizes, deviations show up which can be characterized by a Pareto scale parameter $s_0$ that goes to zero in a linear manner as the current approaches the critical value $I_{\rm p}$ (left-hand graphs and insets in Figure \ref{fig:avalanchestat}). Differences between D-HFN and S-HFN concern only the numerical values of $\tau$ and $s_0$, which are both smaller for the S-HFN but approach common values at failure. The behavior of the R-HFN is qualitatively different from the hierarchical networks but identical to that of a reference network with completely random cross links. In both cases, one finds the same truncated power law distributions with exponent $\tau = 2.2$ and a cut-off that
diverges as the current approaches $I_{\rm p}$. Since the R-HFN has the same distribution of cross-link lengths as the D-HFN but the same exponential distribution  of gap sizes as the random reference network, we can safely conclude that the robust scale-free behavior of the avalanche statistics in the hierarchical networks results from the scale free gap size distribution. This expectation is in line with the fracture pattern of a D-HFN in Fig. \ref{fig:HFN}, bottom right, which demonstrates that the final crack is deflected on all scales by the 
vertical gaps which interrupt stress transmission at the crack tip, leading to a super-rough crack morphology. This qualitative idea is borne out by a quantitative analysis of the distribution of vertical deflections $\Delta y$ of the crack which is characterized by truncated power laws, $p(\Delta y) \propto \Delta y^{-\theta} \phi(\Delta y/L)$ where the cut-off is given by the system size. The observed exponent $\theta = 1.75$ differs from the value $\theta' = 2$ for the gap size distribution along a horizontal line, indicating non-trivial dynamics as stress concentrations at the tip of the emergent crack interact with the network morphology. R-RFN and R-HFN, on the other hand, exhibit an exponential distribution of $\Delta y$ with an average deflection that is slightly larger than the mean gap size.
\begin{figure}
\begin{center}
\vspace{-0cm}
\includegraphics[width=8cm]{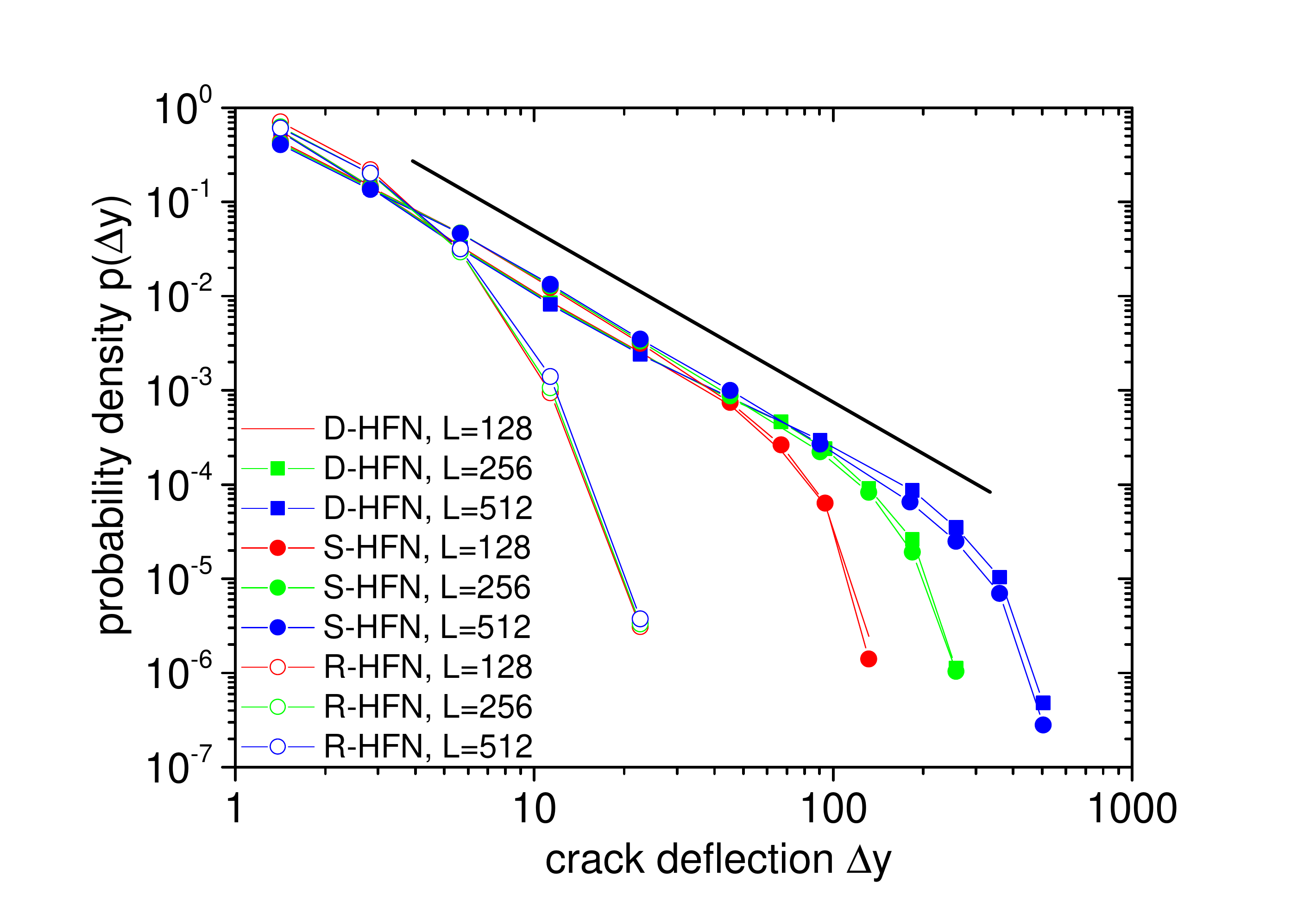}
\caption{Distributions $p(\Delta y)$ of crack deflections in the load-parallel direction for deterministic, shuffled and rotated HFNs of 
sizes $L=128$, $L = 256$ and $L=512$: the straight line represents a power law of exponent $\theta = 1.75$.}
\label{fig:cracks}
\end{center}
\end{figure}

We have proposed a simple model of stress redistribution and failure in a model material with a hierarchical microstructure. Analogously to heterogeneous materials that lack multi-layer hierarchical organization, damage accumulation proceeds intermittently in the form of avalanches, which are broadly distributed in size. We observe however that in the hierarchical case this phenomenology cannot be interpreted as critical behavior in the vicinity of a continuous phase transition, as paradigmatically implemented in fiber bundle models with equal load sharing. Avalanches with power-law distributions without apparent cut-off are observed \textit{generically}, i.e. for any value of the applied load. Avalanche exponents vary continuously, suggesting that the concept of universality class cannot be invoked. We argue that failure patterns, as well as deformation/load patterns, arise naturally from the hierarchical microstructure of the deforming medium, which is scale invariant by construction. The fracture patterns reflect the same scale invariance and strongly differ from the self-affine crack morphologies
generally observed in non-hierarchical random fuse networks \cite{Zapperi2005_PRE}. It remains to be investigated whether the same is true for the basic mode of fracture, which is controlled by nucleation-and-propagation of a critical crack in non-hierarchical RFN whereas Figure  
\ref{fig:HFN}, bottom right, suggests a failure mode by coalescence of multiple flaws as propagation of nucleated cracks is interrupted by the presence of hierarchically distributed gaps. Further work is needed to systematically quantify how the scale free dynamics of damage accumulation and the ensuing crack profiles relate to the parameters governing the scale-free microstructures (exponents of the distribution of link and gap sizes), which can be 'tuned' by changing the number of horizontal and vertical links in the D-HFN generator. Further work is also needed to clarify to which extent the hierarchical microstructure yields substantial benefits in terms of overall toughness and flaw tolerance, and how the 'optimal' microstructure morphology for a network of unreliable components may look like. 


{\bf Acknowledgements} We acknowledge funding by DFG under grants No. Za 171-9 and Mo 3049/1-1. M.Z. also acknowledges support by the Chinese State Administration of Foreign Expert Affairs under Grant No MS2016XNJT044.
  
\bibliographystyle{iopart-num}
\bibliography{BibliographyHFN}

\end{document}